\documentclass[a4paper]{spie}  
\usepackage[]{graphicx}
\def\gsim{\;\lower.6ex\hbox{$\sim$}\kern-9.5pt\raise.4ex\hbox{$>$}\;}
\def\lsim{\;\lower.6ex\hbox{$\sim$}\kern-9.2pt\raise.4ex\hbox{$<$}\;}

\title{On the use of asymmetric PSF on NIR images of crowded stellar fields} 

\author{Giuliana Fiorentino\supit{a}, Ivan Ferraro\supit{b}, Giacinto
  Iannicola\supit{b}, Giuseppe Bono\supit{b,c}, Matteo
  Monelli\supit{d,e}, Vincenzo Testa\supit{b}, Carmelo
  Arcidiacono\supit{a}, Marco Faccini\supit{b}, Roberto Gilmozzi\supit{f}, Marco Xompero\supit{g}, Runa Briguglio\supit{g}
\skiplinehalf
\supit{a}INAF-Osservatorio Astronomico di Bologna, via Ranzani 1, 40127, Bologna, Italy \\
\supit{b}INAF-Osservatorio Astronomico di Roma, via Frascati 33, Monte Porzio Catone, Rome, Italy\\
\supit{c}Dipartimento di Fisica, Universit\`{a} di Roma Tor Vergata, via della Ricerca Scientifica 1, 00133 Rome, Italy\\
\supit{d}Instituto de Astrof\'{i}sica de Canarias, Calle Via Lactea, E38200 La Laguna, Tenerife, Spain\\
\supit{e}Departamento de Astrof\'{i}sica, Universidad de La Laguna, Tenerife, Spain\\
\supit{f}European Southern Observatory, Karl-Schwarzschild-Str. 2, 85748 Garching bei Munchen, Germany \\
\supit{g}INAF-Osservatorio Astronomico di Arcetri, Largo Enrico Fermi, 5, 50125 Firenze, Italy \\
}

\authorinfo{Further author information: (Send correspondence to Giuliana Fiorentino)\\Giuliana Fiorentino: E-mail: giuliana.fiorentino@oabo.inaf.it, Telephone: +39 051 209 53 18}

\begin{document} 
  \maketitle 

\begin{abstract}
 
We present data collected using the camera PISCES coupled with the
Firt Light Adaptive Optics (FLAO) mounted at the Large Binocular
Telescope (LBT). The
images were collected for two different pointings by using two natural
guide stars with an apparent magnitude of R$\lsim$13 mag. During these
observations the seeing was on average $\sim$0.9
arcsec. The AO performed very well, in fact the images display a mean
FWHM of 0.05 arcsec and of 0.06 arcsec in the J-- and in the Ks--band,
respectively. The Strehl ratio on the quoted images reaches 13--30\%
(J) and 50--65\% (Ks), in the off and in the central pointings respectively. On the basis of this sample we have
reached a J--band limiting magnitude of $\sim$22.5 mag and the deepest
Ks--band limiting magnitude ever obtained in a crowded
stellar field: Ks $\sim$23 mag.  

J--band images display a complex change in the shape of the PSF when moving at larger radial 
distances from the natural guide star. In particular, the stellar images become more 
elongated in approaching the corners of the J-band images whereas the
Ks--band images are more uniform. We discuss in detail the strategy used to perform accurate and deep
photometry in these very challenging images. In particular we will
focus our attention on the use of an updated version
of ROMAFOT based on asymmetric and analytical Point Spread Functions. 

The quality of the photometry allowed us to properly identify a feature that clearly shows up in NIR bands: the
main sequence knee (MSK). The MSK is independent of the evolutionary
age, therefore the difference in magnitude with the canonical clock
to constrain the cluster age, the main sequence turn off (MSTO),
provides an estimate of the absolute age of the cluster. The key
advantage of this new approach is that the error decreases by a factor
of two when compared with the classical one. Combining ground--based
Ks with space F606W photometry, we estimate the absolute age of M15 to
be 13.70$\pm$ 0.80 Gyr.    
\end{abstract}



\section{INTRODUCTION}\label{intro}
One of the main unsolved question in astronomy is how galaxies form
and evolve. Resolved Stellar Population (RSP) are one of the most
powerful diagnostic to provide solid constraints on the early 
formation and evolution of the Milky Way (MW). In this context 
Galactic Globular clusters (GGCs) play a crucial role, since they 
are the oldest ($\gsim$12Gyr) stellar systems to probe the early 
formation of both the Halo and the Bulge. This means that GGCs are fundamental laboratories to constrain the 
evolutionary properties of low--mass stars, but also for several 
open problems in modern Astrophysics. One of the key advantages in 
using GCs as tracers of the old stellar populations is that they are
ubiquitous in early and late galaxies and in massive dwarf galaxies\cite{mcconnachie12}.

Absolute and relative ages of GCs together with their proper motions (PMs), 
galactocentric distances and radial velocities are required to properly 
understand the MW formation\cite{marin09,leaman13}. 
The GC absolute ages also provide a lower limit on the epoch of pristine galaxy 
formation, and therefore, on the age of the Universe. 
The measurements of the above diagnostics do require: 
{\it i}) high--quality multi--band images to perform accurate and precise photometry 
in crowded stellar fields; {\it ii}) high--resolution spectra collected with multi-object spectrographs 
to estimate heavy element (iron, $\alpha$, CNO) abundances. The photometry 
allows us to construct Color-Magnitude Diagrams (CMDs) and Luminosity Functions 
(LFs) to evaluate the cluster age and the proper motion. The spectroscopy 
provides the chemical enrichment history and the kinematics. 

During the last twenty years the unprecedented image quality (stable Point 
Spread Function, PSF) and spatial resolution of the Hubble Space Telescope 
(HST) have been a quantum jump in the analysis of stellar populations in 
crowded stellar fields (Galactic Center, Bulge, innermost regions of GCs). 
The consequence is that HST optical images allow us to build CMDs with 
accurate and deep photometry at least five magnitudes fainter than the 
main sequence turn--off (MSTO). The same outcome does not apply to 
near-infrared (NIR) images collected with HST (e.g., with the IR channel of
the WFC3). The diffraction 
limit of HST in the Ks band is $\sim$0.25'', while it is $\sim$0.05'' in 
the I-band. This means limited photometric accuracy in crowded stellar 
fields (center of GCs) and in stellar regions affected either by large and differential
reddening (Galactic center, bulge).

Detailed investigations of the quoted targets do require modern 
NIR detectors at large ground-based telescopes assisted with 
Adaptive Optic (AO) systems. The AO systems can allow us to 
alleviate the effects of the Earth's atmosphere and to reach 
their diffraction limit (e.g., EPICS\cite{kasper10}). Note that this means 
Ks--band images with a FWHM of $\sim$0.07'' in a 8--meter telescope.   
It goes without saying that ground--based observing facilities approaching 
the above limit will have a substantial impact on a very broad range of 
astrophysical problems ranging from extrasolar planets, to stellar 
populations, to the detection of primeval galaxies.
 
These are the main reasons why the astronomical community undertook 
a paramount effort in improving the efficiency and the robustness 
of AO systems. During the last few years become available different 
flavors of AO systems: e.g. Ground Layer Adaptive Optics 
(GLAO) to Single Conjugate Adaptive Optics (SCAO; one 
deformable mirror) 
to Multi Conjugate Adaptive Optics (MCAO; several deformable mirrors) and to Multi-Object Adaptive Optics (MOAO). These 
pioneering AO systems not only demonstrated on sky the feasibility, 
but also provided data to address several astrophysical problems. 
Moreover, and even more importantly, they paved the road for the 
next generation of Extremely Large Telescopes (ELTs\cite{deep11,greggio12,schreiber13}).    
  
The key advantage of the MCAO systems is their uniform correction across a
quite large field of view (FOV$\sim$2x2'), while the SCAO systems ``only'' 
correct a small FOV of $\sim$30x30''. The improvement is impressive as 
demonstrated on sky by the Multi--conjugate Adaptive optics 
Demonstrator (MAD) temporarily available at 
VLT \cite{marchetti08} (see Fig. 1 in Marchetti et al. 2008) and 
more recently by GeMS (Turri et al. 2014, this SPIE conference) 
available at the Gemini South Telescope. MAD had a limited sky-coverage, since the 
closure of the loop required an asterism of three stars brighter than 
V$\sim$13 mag over two arcmin on the sky.
However, MAD was a very successful experiment, since delivered NIR images 
with a uniform PSF across the corrected FOV and produced 
several interesting investigations on RSP  
\cite{momany08,gullieuszik08,moretti09,bono10b,fiorentino11}. 
The accuracy and precision of the CMDs based on MAD images is 
similar or even better than CMDs based on images collected with 
optical and NIR images collected with HST. 
Disentangling two different stellar populations in the bulge GGC
Terzan 5, MAD observations suggested that this cluster is a candidate
building block of the Galactic bulge\cite{ferraro09a}. 
However, the complex variation of the PSF across the FOV and in time 
makes the analysis of MCAO data quite challenging. 

Equipped with one deformable secondary mirror and a high-order pyramid 
wave-front sensor, each LBT eye can work
with a SCAO system that coupled with the high spatial resolution of
PISCES (pixel scale = 0.0193'') can deliver high quality NIR images 
of crowded stellar fields. When first mounted on one of the LBT’s 
(early 2010), the adaptive optics system--FLAO\cite{esposito11}--provided the best NIR images ever collected 
with a ground--based telescope. The new systems succeeded in delivering 
images with a PSF across the FOV that was more than three times sharper 
than images collected with HST. Moreover, during the initial testing 
phase, the LBT’s adaptive optics system was able to achieve 
and unprecedented Strehl Ratio in the Ks--band: from 60 to 80 percent. 
This means an improvement of two-third in image sharpness when 
compared with similar AO systems available at 8-m class telescopes 
\cite{esposito11}.

Current generation of AO systems are a fundamental playground not only 
for the technological challenges, but also to sharpen our fingernail in
analyzing NIR images of crowded stellar fields that will be collected 
with MCAO systems. This will offer the unique opportunity to
revolutionize our approach to Galaxy evolution and to pave the way
for future ELTs (diameter$\ge$30m\cite{deep11,greggio12,schreiber13}). It is of
critical importance for the full exploitation of ELT facilities that
NIR RSP are explored and successfully modeled. AO deep NIR data will
provide crucial inputs for the modeling of cool stars, mainly emitting
in NIR, for which uncertainties related to the effective temperature
evaluation and to the colour-temperature transformations require further
investigation\cite{allard12,allard13}. 

Moreover, recent deep NIR studies indicate a new robust method to estimate 
the age of GGCs using the signature of collisional induced absorption of
molecular hydrogen in low-mass (M$\sim$0.4 M$_{\odot}$) 
MS stars\cite{bono10b} (see Fig. 2 in Bono et al. 2010). This opacity 
mechanism shows up as a well--defined Knee observable in the faint MS 
(MSK), and its position is, at fixed chemical composition, independent of
the cluster age (Mk$\sim$5 and Mj-Mk$\sim$0.7mag). The magnitude
difference between the MSK and the MSTO is thus a solid measurement of
the absolute age and, by definition, is independent of reddening and
distance. When the MSTO is instead anchored to Horizontal Branch
stars, their age dependence is a major systematic on the final age
estimation. Ground based telescopes equipped with SCAO or MCAO are 
expected to detect and calibrate the MSK over a
broad metallicity range, providing absolute ages with high
accuracy ($\sigma$(MSK--MSTO)$\sim$ $\sigma$(MSTO)/2$\sim$ 1Gyr). 

In this paper we focus our attention on M15 which belongs to the
metal--poor tail of the GGC distribution ([Fe/H $\lsim$-2 dex]). 
Moreover, M15 appears to be in an advanced dynamical state 
(post core collapsed), this means that the innermost cluster regions 
display a well defined sharp peak in luminosity, and in turn in density 
profile\cite{king71}. It has also been suggested that M15 harbors 
an Intermediate Mass Black Hole\cite{vandermarel02}. The above feature made M15 
a fundamental laboratory to test the accuracy and the precision of the 
approach adopted to perform photometry in crowded stellar fields (ALLFRAME\cite{stetson94}).

   \begin{figure}
   \begin{center}
   \begin{tabular}{c}
   \includegraphics[height=10.cm]{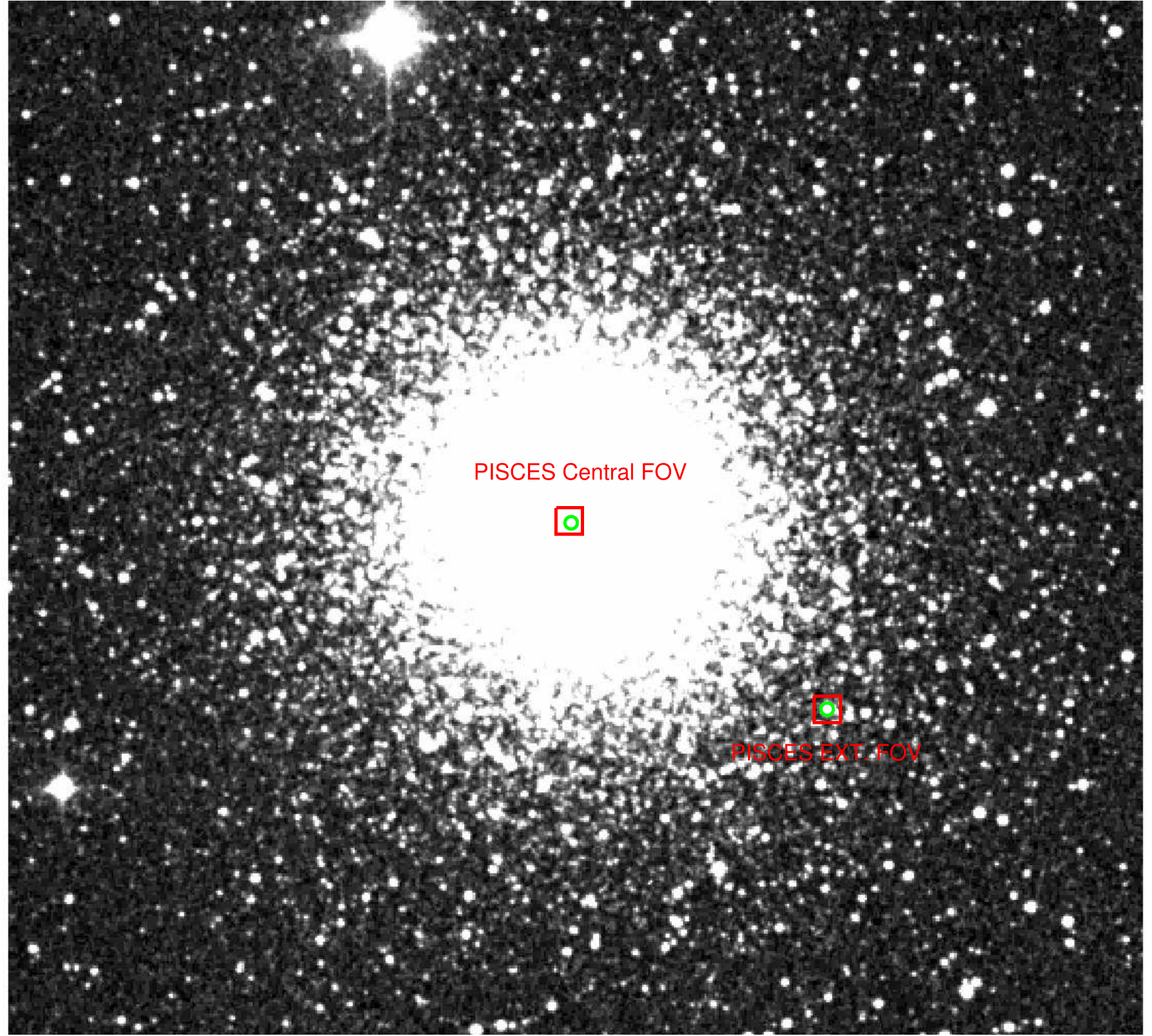}
   \end{tabular}
   \end{center}
   \caption[example] 
   { \label{fig1} 
  Finding chart of M15 showing the two PISCES pointings (red squares) 
with a FOV of 21$\times$21''. The natural guide stars adopted to close 
the loop of the AO system are marked with green dots.}
   \end{figure}

   \begin{figure*}
   \begin{center}
   \begin{tabular}{c}
   \includegraphics[height=6.cm]{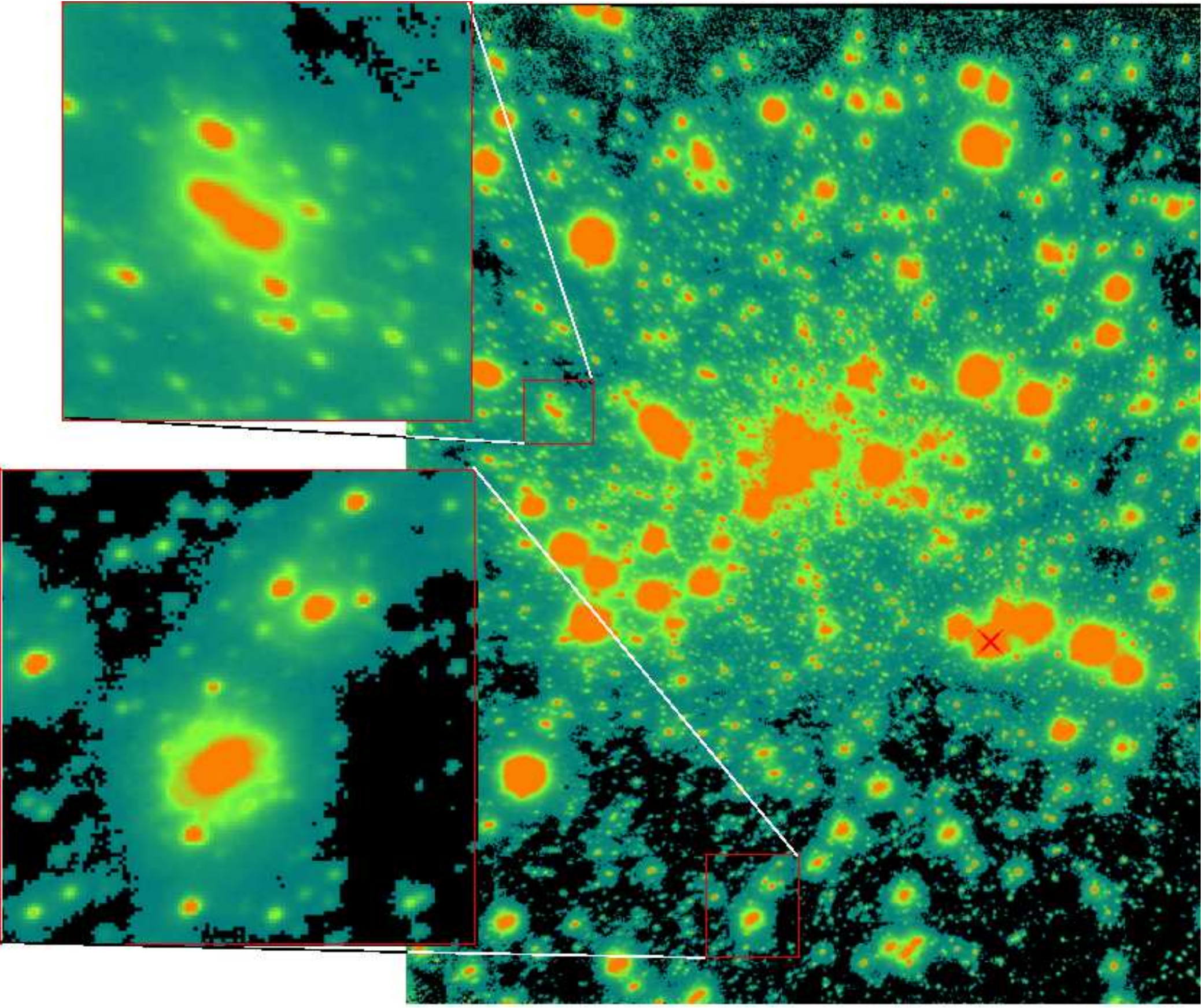}
   \includegraphics[height=6.cm]{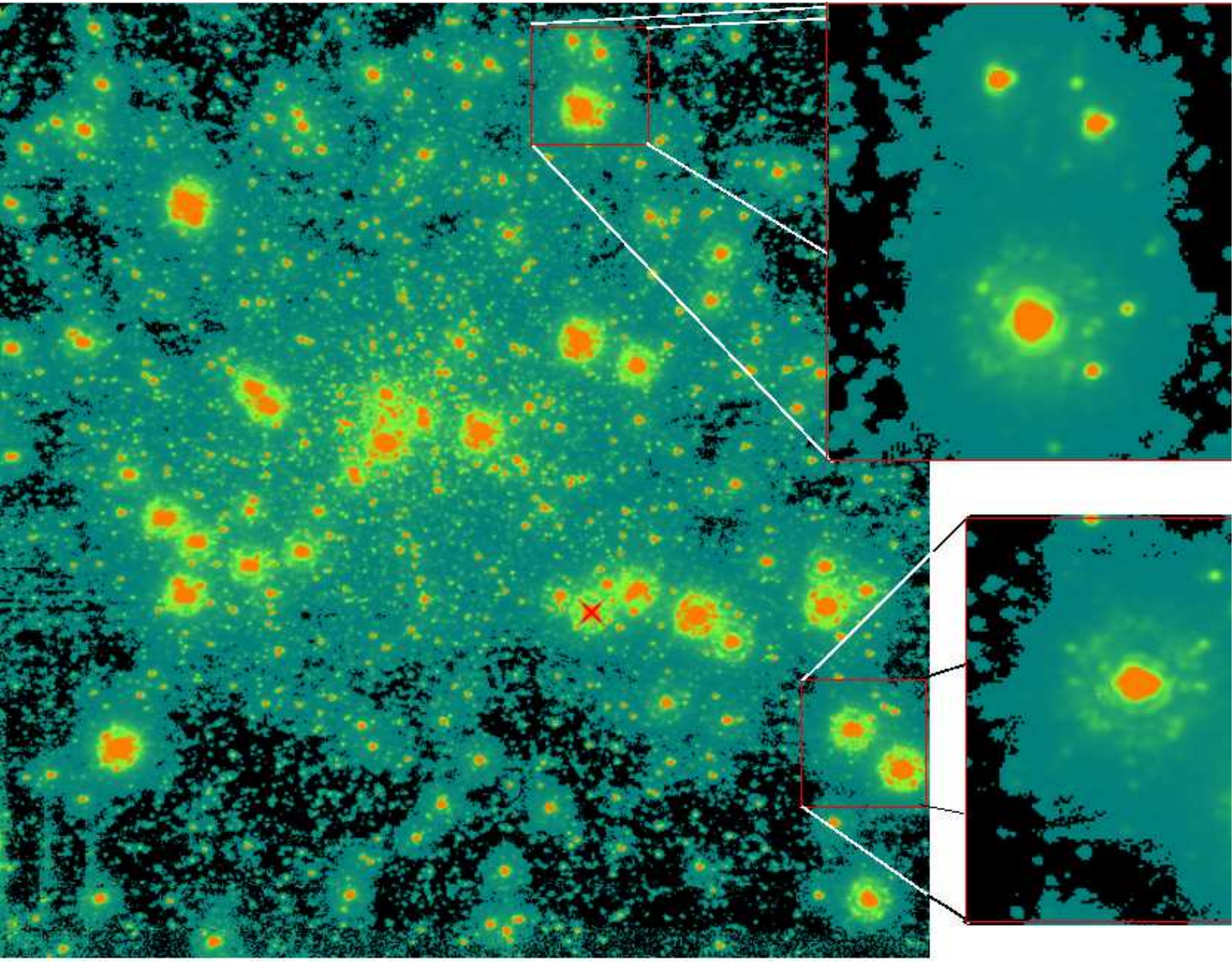}
   \end{tabular}
   \end{center}
   \caption[example] 
   { \label{fig2} 
 Left: J-band image collected with PISCES@LBT ($t_{exp}$=10 s, see Table~\ref{tab1}). 
The two small images located on the left side display the zoom of two regions 
located at different distances from the NGS highlighted with a red
cross in the image. 
The difference in the shape of the stars between the two regions clearly 
show a difference in the perfomance of the AO system. 
Right: Same as the left, but for a Ks--band image ($t_{exp}$=15 s) . 
The Ks--band image shows a better correction than J-band image, and indeed 
the shape of the stars is more regular and circular across the FOV. 
The speckles generated by the AO correction are present both in the J and in 
the Ks--band images, the extended seeing halos can also be easily identified.}
   \end{figure*}

\section{Photometric data} 

The NIR images adopted in this investigation were secured in 
October 2011 with PISCES by using the DX (right) 8.4 m telescope 
of LBT where the FLAO was originally mounted. This AO system relies 
on two key components, namely an adaptive secondary mirror 
with 672 actuators and an innovative high-order pyramid 
wave-front sensor.

The scientific target was the Galactic Globular M15 (RA$=$21:29:58.33, DEC$=$+12:10:01.2) 
visible for more than about four hours with an air mass smaller than 1.3. 
Data were acquired for two different different pointings, the former 
is centrally located, while the latter is located at 3 arcmin form 
the center in the South-West direction. 
Two natural guide stars adopted to close the loop of the AO system are: 
NGS1, RA= 21:29:58.616, DEC=+12:09:56.34, R=12.6 mag central pointing 
and NGS2, RA= 21:29:44.6457, DEC=+12:07:30.697, R=12.9 mag off--center 
pointing.  One of the key advantages in observing Galactic 
Globulars is that NGSs can be easily identified in the innermost cluster 
regions. Cluster red giant stars are typically brighter than R$\sim$13 mag  
up to distances of 10--16 kpc.

We collected 18 J-band and 20--Ks band images for the central pointing 
with a total exposure time of 108 and 170 s, while for the off--center pointing 
we secured 20 J-band and 42 Ks--band images for a total 600 and 630 s 
(see Table~1). The external DIMM seeing ranges from 0.7 to 1.1 arcsec for the 
central pointing and from 0.6 to 0.95 arcsec for the off--center pointing.
These weather conditions and the selected NGSs allowed us to reach a Strehl ratio up to 65\% in Ks and 30\% in J in the central pointing and up to 50\% in Ks and 13\% in J in the off--center pointing. Fig.~\ref{fig1} shows the position of the pointings together with the position 
of the adopted NGSs.

\begin{table}[h]
\caption{Observing log.} 
\label{tab1}
\begin{center}       
\begin{tabular}{|l|l|l|} 
\hline
\rule[-1ex]{0pt}{3.5ex}  INSTRUMENT & PASSBAND  & N$_{obs} \times$TEXP   \\
\hline
\rule[-1ex]{0pt}{3.5ex}  PISCES-central FOV & J  & 9$\times$2 s     \\
\rule[-1ex]{0pt}{3.5ex}  PISCES-central FOV & J  & 9$\times$10 s     \\
\rule[-1ex]{0pt}{3.5ex}  PISCES-central FOV & Ks  & 10$\times$2 s    \\
\rule[-1ex]{0pt}{3.5ex}  PISCES-central FOV & Ks  & 10$\times$15 s   \\
\rule[-1ex]{0pt}{3.5ex}  PISCES-external FOV & J  & 20$\times$30 s   \\
\rule[-1ex]{0pt}{3.5ex}  PISCES-external FOV & Ks  & 42$\times$15 s   \\
\hline 
\end{tabular}
\end{center}
\end{table} 

In order to provide a more clear idea concerning the AO performance 
that can be reached with PISCES, Fig.~\ref{fig2} shows two images collected 
in the J (left) and in the Ks (right) band for the central pointing. 
A glance at the shape of the stars plotted in these images clearly shows 
that the PSF in J-band become more and more elliptical as a function of 
the distance from the NGS. 
The FWHM increases from 0.04 arcsec in the very center of the image to about 0.07 arcsec in its outskirt. 
This effect is more mitigated in the Ks--band, and indeed the shape of the PSF 
remains almost circular across the entire image (see the zoom on individual 
stars). 
The FWHM is on average $\sim$0.06 arcsec across the entire image.   
The above empirical evidence indicates that stars typically cover about 2$\times$2 and 3$\times$3 pixels in the J and in the 
Ks--band, respectively. 

In passing we note that for the same cluster space images are also available collected using the F160W--band with WFC3/IR at HST. The images overlap with the off--center pointing, the diffraction limit of HST in this filter is $\sim$ 0.17 arcsec and the pixel scale of the IR channel is 0.13 arcsec per pixel.   
Fortunately enough, a good set of optical space images of the central 
regions are also available. They have been collected with ACS/WFC at 
HST in the F814W (similar to I-band) and in the F606W (similar to a 
V-band). The diffraction limit in F606W--band is 0.06 arcsec and the pixel scale of 
the WFC is 0.04 arcsec per pixel. 
The above evidence indicates that ground-based NIR images are 
typically oversampled when compared with optical and NIR space images 
collected with HST.

\section{Data reduction} 

Photometry of crowded stellar fields became a solid opportunity during the late 
eighties thanks to the development of sophisticated automatic techniques to 
perform the photometry on digitalized photographic plate images. The use 
and the role played by these packages became even more crucial in handling 
hundreds of optical images collected with modern CCDs.  
The most commonly used packages to perform PSF fitting photometry in
crowded stellar fields are DAOPHOT (ALLSTAR/ALLFRAME) \cite{stetson87,stetson90}, 
ROMAFOT \cite{buonanno83,buonanno89}, DOPHOT \cite{schechter93}, SExtractor\cite{bertin96,bertin11}, Starfinder \cite{diolaiti00} and software ad hoc developed to deal with HST images \cite{anderson06,anderson10}. The use of these packages made possible the
construction of very deep and precise CMDs based on ground--based 
and space images that allowed us to investigate in detail RSPs in the 
very center of GGCs\cite{bedin09} and in the Bulge\cite{lagioia14}. Note that the use of the PSF photometry in crowded 
stellar fields is mandatory not only to improve the precision of the photometry 
(identification of faint companions), but also to improve the limiting magnitude 
(identification of faint objects). The latter point is even more crucial in 
dealing with NIR images, since they are sky limited.  

   \begin{figure*}
   \begin{center}
   \begin{tabular}{c}
   \includegraphics[height=8.5cm]{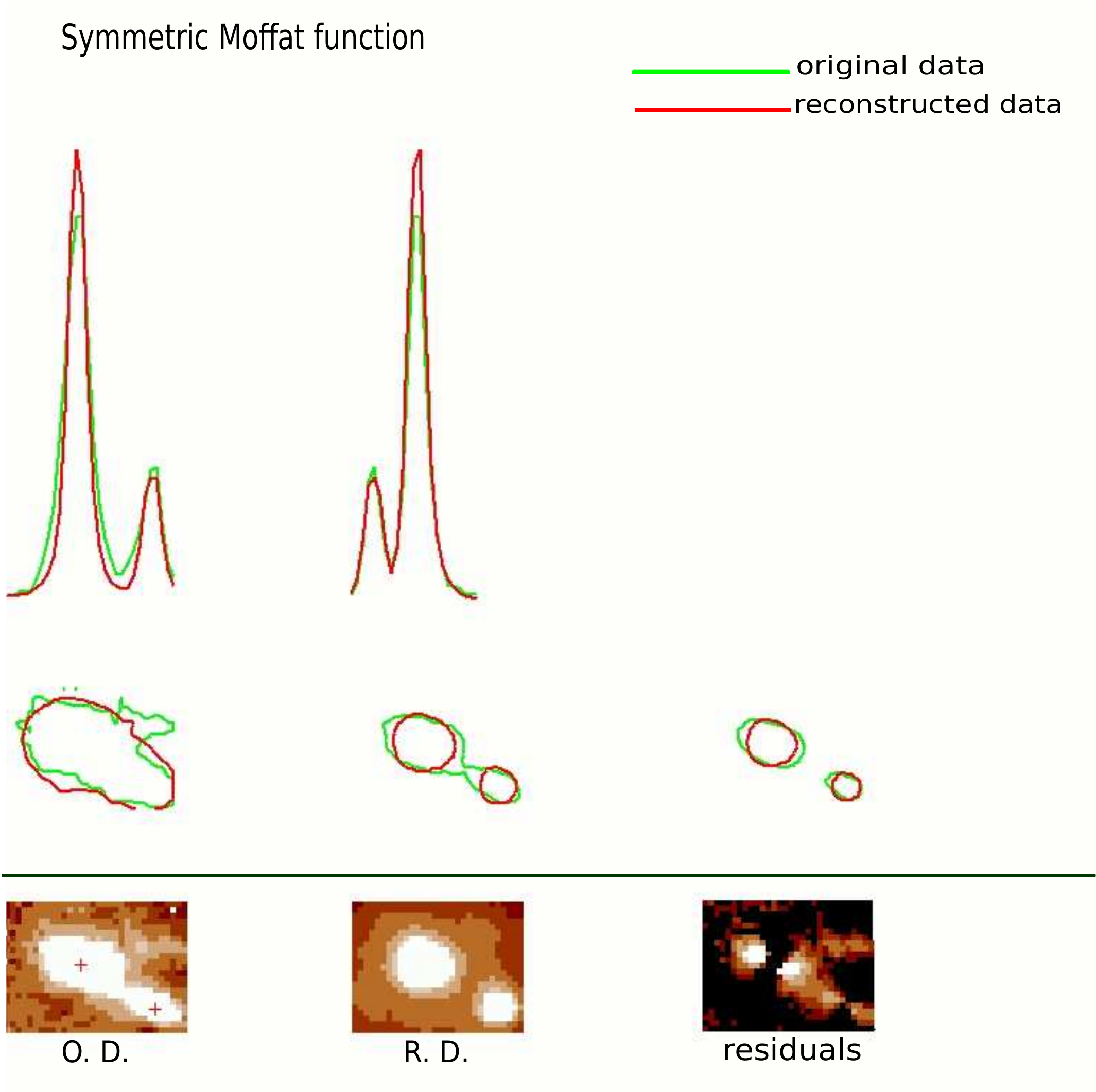}
   \includegraphics[height=8.5cm]{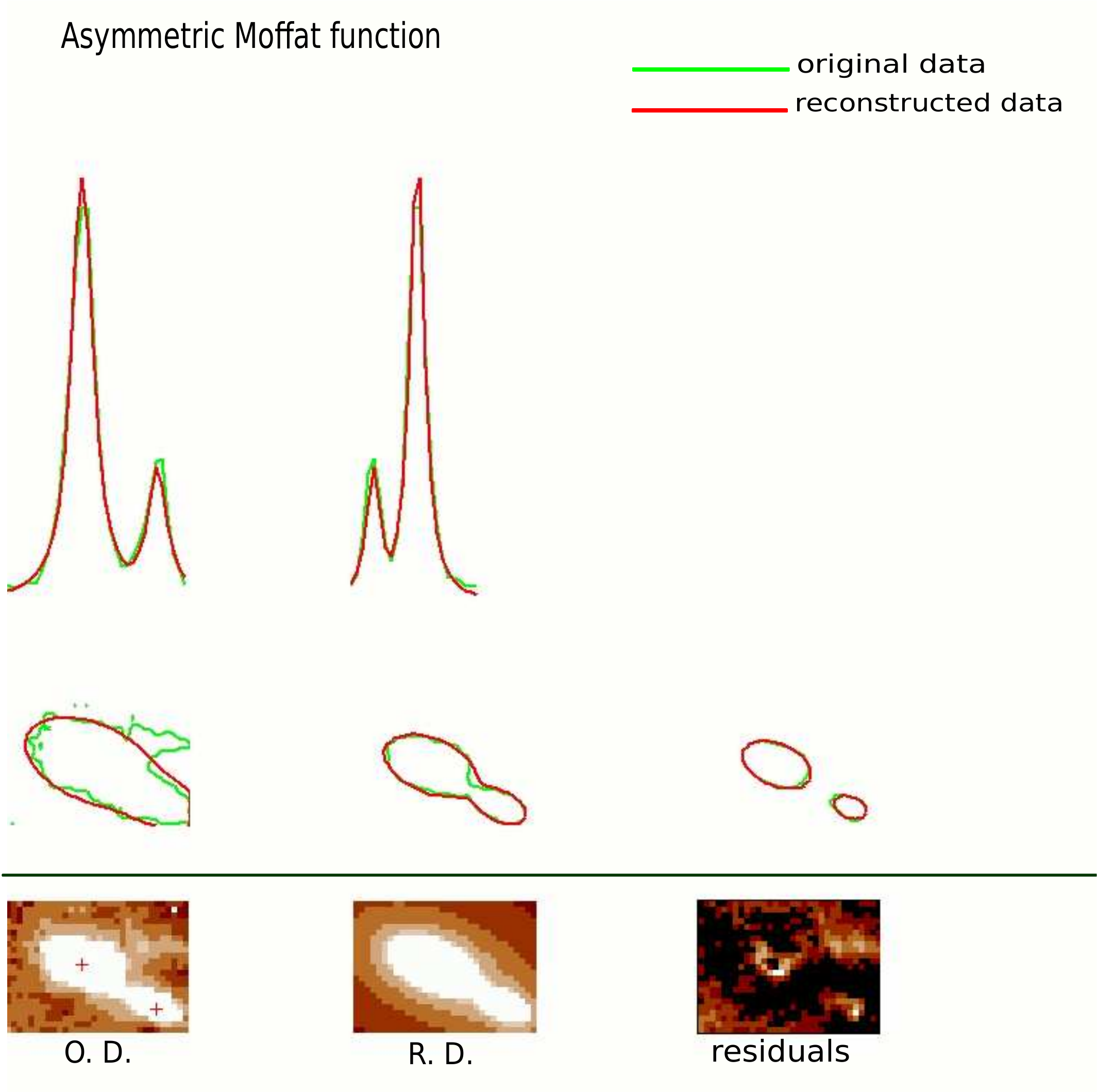}
   \end{tabular}
   \end{center}
   \caption[example] 
   { \label{fig3} 
Left: 2D projection of the analytical PSF fitting on two close 
stars performed with ROMAFOT on a single J-band image. The fit uses 
a classical symmetric Moffat function. The green 
lines shows the original data (O.D.), while the red one the expected 
profile (reconstructed data, R.D.). The left and middle plots on top 
of the figure display the projection along the X- and the Y-axis. 
The left and middle contours in the middle display the view 
from the Z-axis. The left and the middle images in the bottom 
display the original data and the reconstructed data. The right 
contours in the middle and the right image in the bottom display 
the residuals once the deconvolution of the original data was performed.      
Right: Same as the right, but the analytical PSF fitting was performed 
by using an asymmetric Moffat function. The use of the this PSF significantly 
improves the quality of the fitting, and indeed the residuals decrease 
over the entire area covered by the two close stars.}
\end{figure*}

   \begin{figure*}
   \begin{center}
   \begin{tabular}{c}
   \includegraphics[height=8.5cm]{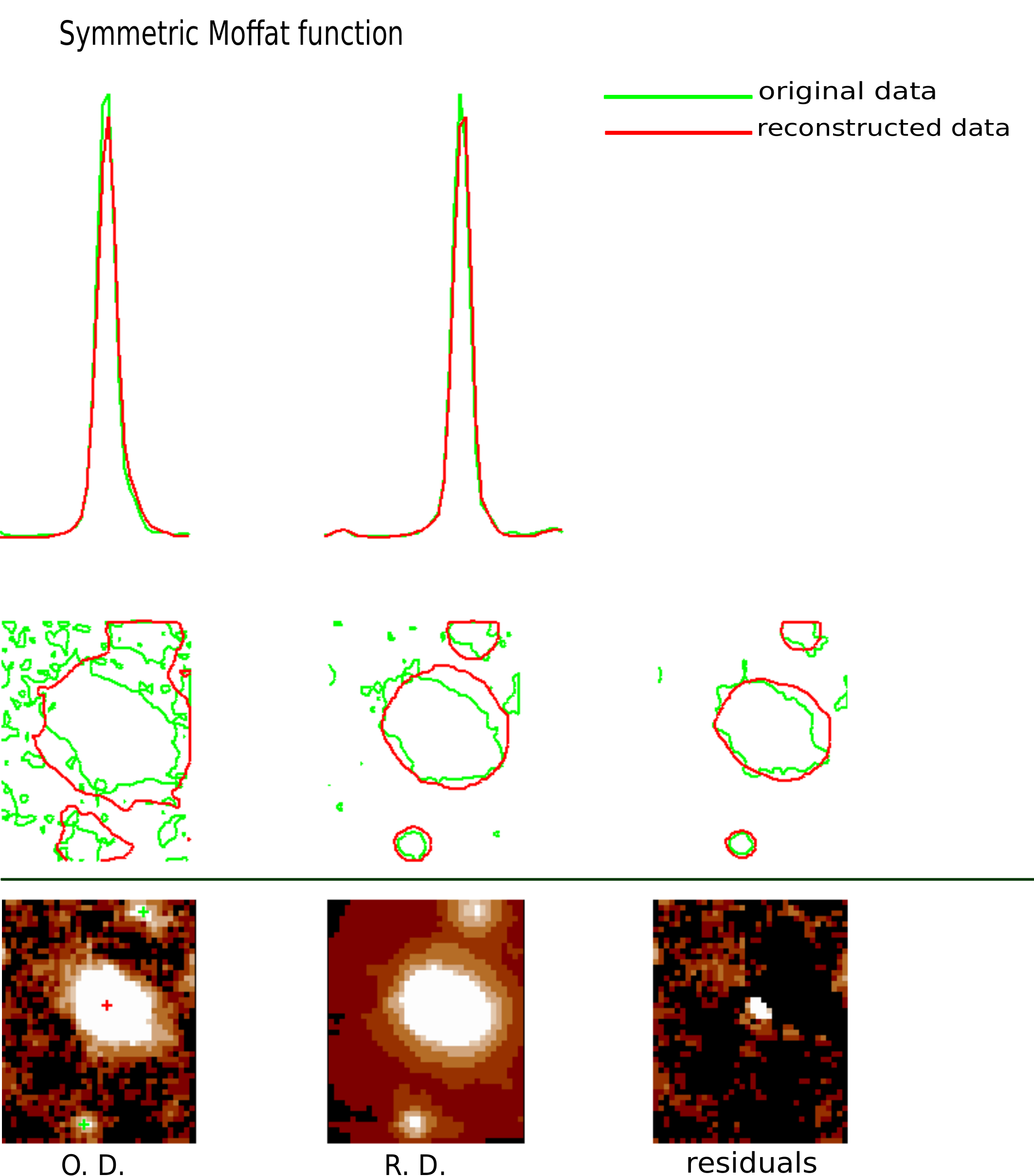}
   \includegraphics[height=8.5cm]{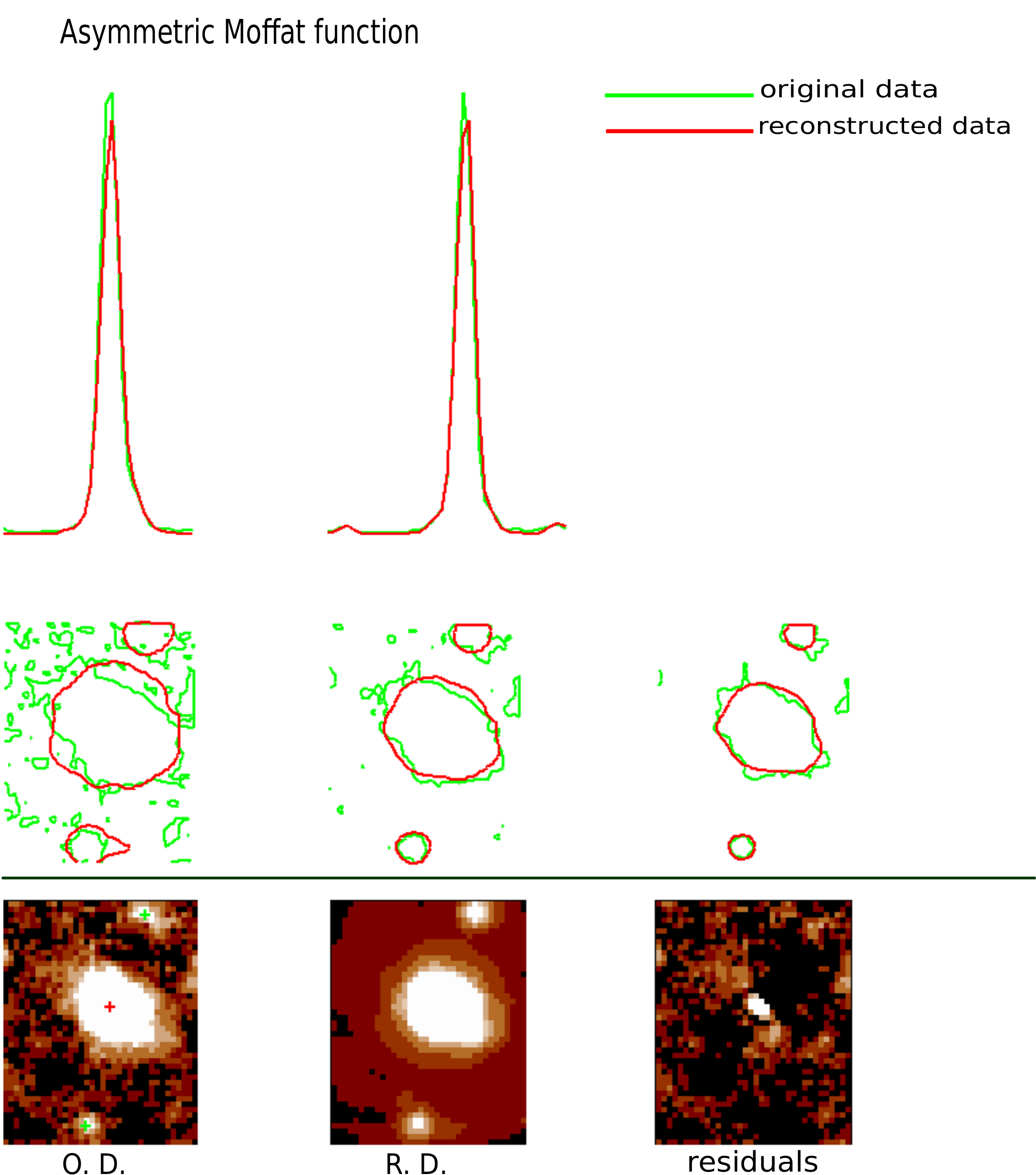}
   \end{tabular}
   \end{center}
   \caption[example] 
   { \label{fig4} 
Same as in Figure ~\ref{fig3}, but for an analytical PSF fitting performed on 
three close stars of a Ks--band image. The asymmetric PSF is still improving 
the quality of the fit, but the improvement is smaller than for the J-band 
image.}
\end{figure*}

The use of ground--based NIR images collected with AO systems seems 
to suggest that a further development of the packages currently available 
is required. The shorter wavelength images (tipically J--band) display a complex change of 
the shape of the PSF across the FOV. This evidence further support the 
suggestion that the performance of a AO system does depend on the Strehl 
ratio, but also on the image quality and on the temporal stability of 
the PSF. In the following we discuss in more detail while the latter 
parameters play a crucial role in the photometry of crowded stellar 
fields. Note that there is mounting evidence that NIR images collected 
with MCAO systems (MAD at VLT, \cite{marchetti08}) have a more uniform 
and stable PSF across the FOV. This is the reason why in dealing with 
these images have been adopted classical approaches\cite{bono10b,fiorentino11}.  

In dealing with NIR images collected with SCAO systems 
(NACO at VLT, PISCES at LBT) we decided to use ROMAFOT. The key 
advantage in using ROMAFOT is that we can check the different steps 
(fit, map of the residuals) of the individual PSF fitting thanks to 
the graphical interface. The main drawback is that it is lengthy.
The key advantages in using analytical PSF fitting of crowded stellar 
fields when compared with aperture photometry and with numerical 
PSF\cite{stetson87} have been widely discussed in the literature.   
We mention that the analytical PSF improves at fixed magnitude 
the photometric precision thanks to the identification of faint 
companions and the opportunity to discriminate between stars and 
galaxies. Moreover, it improves, at fixed exposure time, the 
limiting magnitude thanks to the identification of faint stellar 
sources. The latter point is even more crucial in dealing with 
NIR images, since they are sky limited.    

In the following we use a Moffat function that can be expressed as 
  
{\small
\begin{equation}
PSF(h, x_0,y_0, \sigma, \beta, x,y)=h*(1+\frac{(x-x_0)^2+(y-y_0)^2}{\sigma^2})^{-\beta} 
\end{equation} 
}
The $\beta$ parameter fixes the slope of the wings, larger is the value steeper is 
the slope. We adopted this analytical PSF because it provides a good description 
of the stellar profiles both in seeing limited and in diffraction limited 
images. In the ROMAFOT environment the analytical PSF fitting is performed 
simultaneously with an assumption that accounts for the local sky--background. 
This means that once the shape of the PSF has been fixed
($\beta$ and $\sigma$ values), by using PSF stars, the number
of unknowns for an isolated stellar profile is three
(x$_0$, y$_0$, h) plus the term that accounts for the local sky--background. This is the main reason why accurate and precise
photometry of crowded stellar fields does require that stars cover at least 
2$\times$2 pixels. To overcome this problem the use of 
specific analytical\cite{buonanno83} and/or numerical\cite{stetson87} 
PSFs has been suggested. Fortunately, the new CCD images 
have appropriate pixel scales. 

To perform accurate PSF photometry of stellar images with asymmetric profiles 
have been suggested different approaches. The most simple approach is to split 
the FOV in a number of smaller subfields and to perform the selection of 
PSF stars on these individual subfields. A similar approach relies on the selection 
of a sizable sample of PSF stars across the entire field of view and to assume 
either a quadratic or a cubic change of the PSF across the image. However, 
these approaches become quite complicated in dealing with crowded stellar 
fields, since the number of good isolated PSF stars is limited. The use of 
specific algorithms overcomes some of the quoted problems and appears quite 
useful in dealing with images that are on the verge to be undersampled\cite{buonanno89}.  
It has also been suggested to use a ``Penny'' function i.e. the sum of a 
Gaussian and of a Lorenz function, in which the latter one can be either 
tilted or not tilted when compared with the Gaussian function\cite{stetson87}.

The quoted approaches present several clear advantages, however, after 
different tests and trials on SCAO images we decided to perform the 
analytical PSF fitting by using a new asymmetric Moffat function:\\

{\small
PSF(h, x$_0$, y$_0$, a, b, c, $\theta$, $\beta$, x, y)=\par

\begin{equation}
h*(1+\frac{((x-x_0)cos \theta+(y-y_0)sin \theta)^2}{a^2}+\frac{((y-y_0)cos \theta+(x-x_0) sin \theta)^2  * (c * ((x-x_0)cos \theta+(y-y_0)sin \theta))^2}{b^2})^{-\beta}
\end{equation} 
}

In spite of the apparently complicated analytical representation, the new 
analytical function is relatively simple. The main features are the following:  
\begin{itemize}
\item 1) Wings -- The $\beta$ paramater of the Moffat function is fixed using 
PSF stars and does not change across the image.  
\item 2) Ellipse -- The core of the Moffat function is no more a circle, but an ellipse 
characterized by its semi--major (a) and semi--minor (b) axes. 
\item 3) Tilt -- The ellipse is not aligned with the X and Y axis, but has an inclination 
angle--$\theta$--with the X axis. 
\item 4) Asymmetry -- The 3D shape of stellar images can be asymmetric thanks to 
the $c$ parameter, i.e. this constrains the departure from an ellipsoidal shaped 
volume into an egg shaped volume.
\end{itemize}

The number of unknowns of the new asymmetric Moffat function are eight (plus 
the sky--background value).  The four unknowns affecting the shape of the PSF (a, b, c, $\theta$) 
change across the image and they are fixed by fitting the stellar profile of 
several, possibly isolated, PSF stars across the frame. We derive for every image, 
on the basis of the PSF stars, an almost spatially uniform grid of values for 
the five unknowns. This approach allow us to decrease the number of unknowns, 
and in turn to speed up the PSF fitting of both isolated and blended stars.  
The new ROMAFOT according to the position of the star in the grid performs 
a quadratic interpolation of the four shape parameters across 
the grid and then performs the fit of individual stellar profile. This means 
that the unknowns of the new approach are still three ($h$, the height;
$x_0$, $y_0$, the centroid) plus the sky--background value. 
The reasons why we devised the above reduction strategy are the following:
1) Current ground-based NIR images collected with AO systems are typically 
oversampled. Indeed, the sampling parameter, defined as FWHM/Pixel scale, 
is of the order of $\sim$2.5--3 for both J and Ks--band 
images\footnote{Note that the typical FWHM for J- and Ks--band images is 
0.05 and 0.06 arcsec, while the PISCES pixel--scale is 0.0193''/pixel.}  
This means that the deconvolution of isolated stars does allow us 
to provide solid measurements of the unknowns affecting the shape of the PSF. 
2) We performed series of numerical experiments on synthetic images and we 
found that the asymmetrical analytical PSF with four unknowns and five parameters 
fixed {\em a priori} works very well across the entire image. In this context 
it is worth mentioning that the fit of individual stellar profiles with an 
asymmetric PSF (nine unknowns) provides less accurate deconvolutions of 
blended stars in the innermost cluster regions and of fainter stars.  
3) The numerical complexity of the PSF fitting significantly decreases.

   \begin{figure*}
   \begin{center}
   \begin{tabular}{c}
   \includegraphics[height=8.5cm]{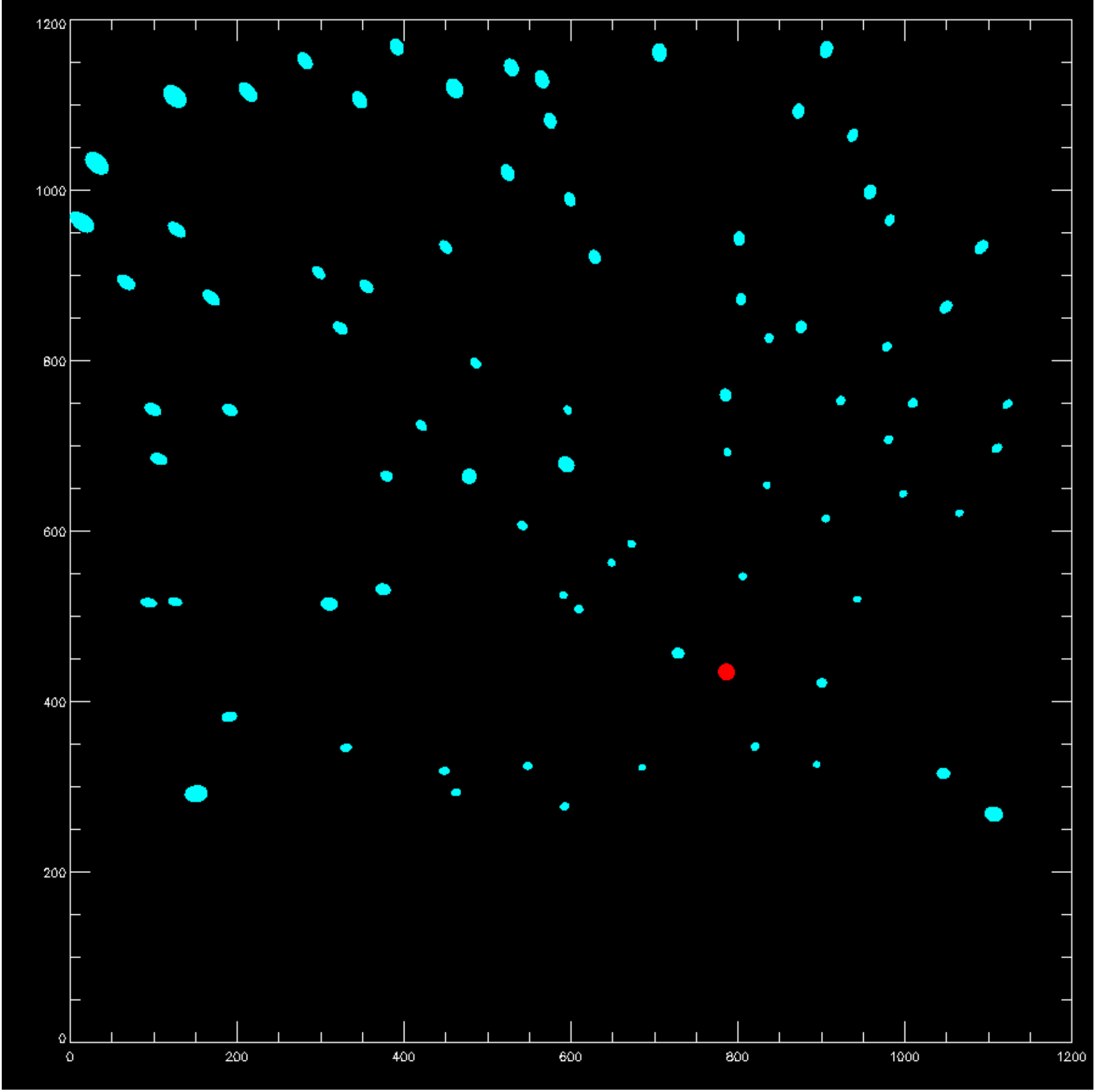}
   \includegraphics[height=8.5cm]{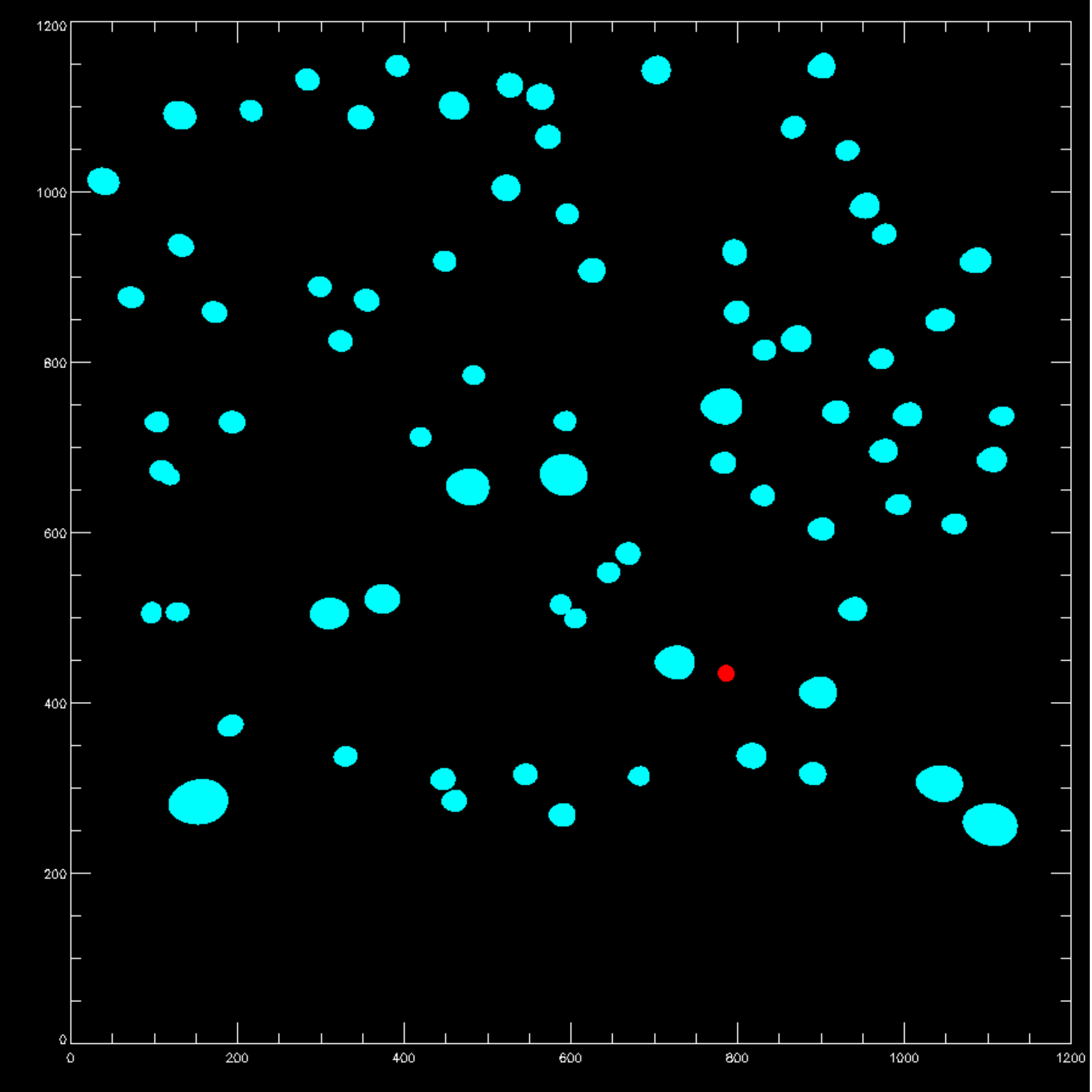}
   \end{tabular}
   \end{center}
   \caption[example] 
{Synthetic J- (left) and Ks--band (right) images of the adopted 
PSF stars based on the new asymmetric Moffat function (see eq.~2.). 
The location of the NGS is marked in red.\label{fig5}. 
}
\end{figure*}

In order to provide a more quantitative analysis of the difference 
between symmetric and asymmetric analytical PSF 
Figs.~\ref{fig3}--\ref{fig4} display the details of the fit. 
The plots on the left display the Original Data (OD, green) and the 
reconstructed Data (RD, red) of the stellar profile projected onto 
the X- (left) and onto the Y-axis (middle) by using a symmetric 
Moffat function. The comparison shows that the 
analytical symmetric PSF does not properly fit the peak of the observed 
profile. The contours at three arbitrary cuts plotted in the middle 
panels display a similar problem, but in the wings of the profiles (see also the images 
plotted in the bottom panel). The discrepancy between observed and 
reconstructed profile is even more clear in the residual map plotted 
in the right bottom panel.     

The comparison between observed and reconstructed profile based on an 
analytical asymmetric PSF is showed in the right panels of Fig.~4. 
Data plotted in these panels display that the new fits takes account 
of both the peak and the wings. Indeed the map of the residuals attains 
vanishing values over the entire area covered by the two close stars.   
Fig~\ref{fig4} shows the same comparison, but for three nearby stars 
on Ks--band image. The use of the asymmetric  analytical PSF still 
improves the quality of the fit, but the difference is less clear 
than for the J-band.  

In order to constrain on a quantitative basis the difference between 
the fit performed with the symmetric and the asymmetric Moffat Function, 
we estimated the difference in flux, over an area of 10$\times$10 pixels 
around the center of the star, between the reconstructed and the original 
data. In particular, we adopted the following formula for the residuals:

{\small
\begin{equation}
RESIDUAL~~FLUX= N* \sqrt{(\sum((O.D._i-R.D._i)^2)/N)}/\sum(O.D._i-background)*100
\end{equation}
}
We found that the residuals in the J-band when using the asymmetric and the 
symmetric Moffat function are $\sim$16\% and $\sim$34\%, respectively. This means 
a decrease of at least a factor of two. The difference in Ks band is significantly 
smaller and indeed the two residuals attain similar values, i.e. 20\%
and 22\%, respectively.

The use of the new asymmetric analytical PSF fitting allowed us to perform 
accurate photometry across both J- and Ks--band images. In particular, the 
quality of the individual fits makes the final check of the residual map 
useless, since the local shape of the PSF takes properly into account 
the radial variations. To constrain on a more quantitative basis the 
impact of the new asymmetric PSF, we reconstructed synthetic J- (left 
panel of Fig.~5) and Ks--band (right panel of Fig.~5) images to be compared 
with real NIR images collected with SCAO systems. The elongated shape 
(eggy--like)\footnote{The semi--major axis in the J-band ranges from 
0.7 to 2.9 pixels, while in the Ks--band from 2.6 to 5.9 pixels.  
The semi-minor axis attains similar values, and indeed it ranges from 
0.6 to 1.9 pixels (J-band) and from 2.3 to 4.4  (Ks--band).}
and the orientation of the stellar images of the left panel 
clearly display the same radial trend from the NGS (red dot) showed 
by real images (see Fig.~\ref{fig2}). The similarity also applies to stars 
located in the corners of the image. The mild change in the shape of 
the stellar images can also be easily identified in the Ks--band image, 
but it is less evident when compared with the J-band.

  \begin{figure*}
   \begin{center}
   \begin{tabular}{c}
   \includegraphics[height=8cm]{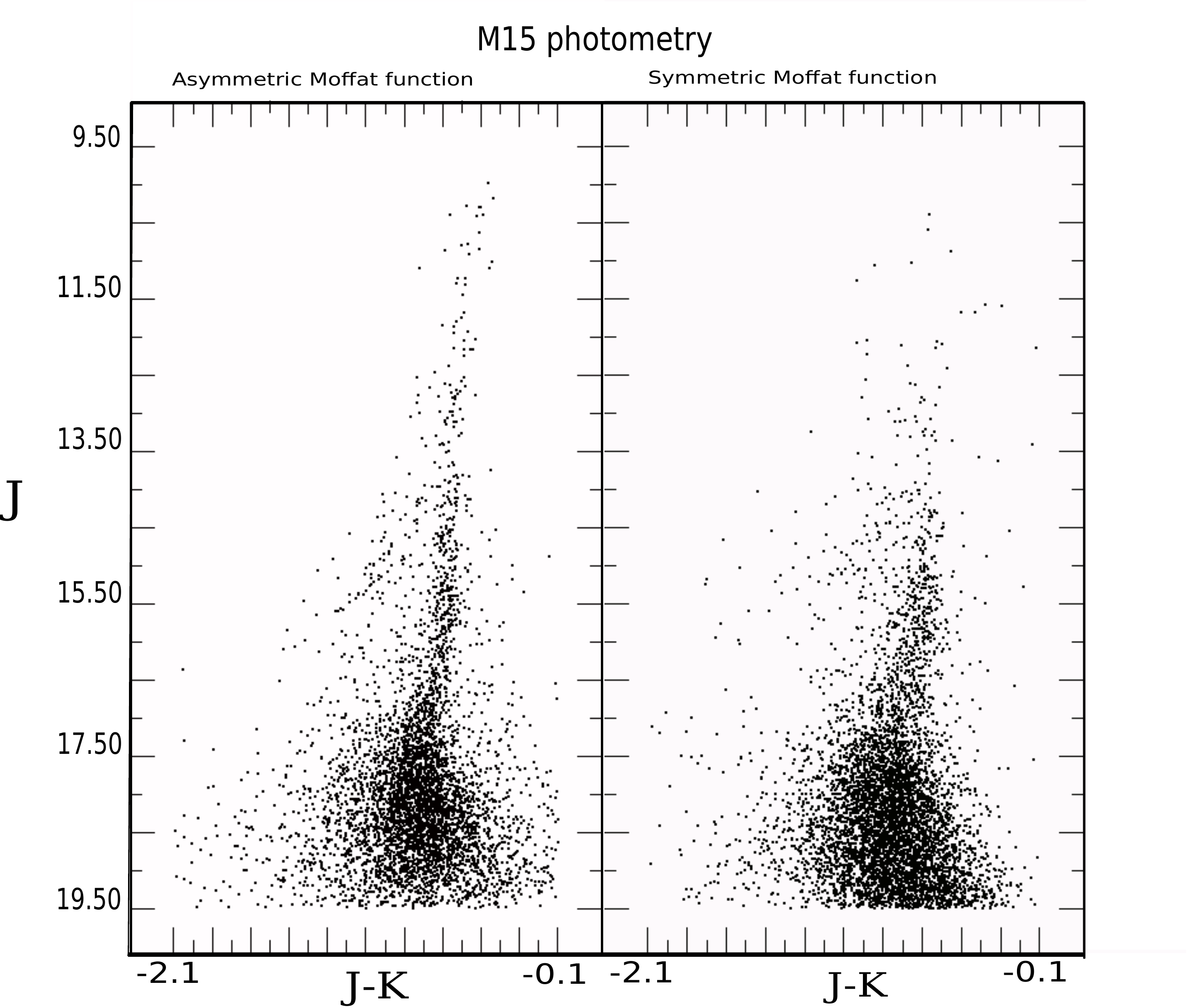}
   \end{tabular}
   \end{center}
   \caption[example] 
   { \label{fig6} 
Left: NIR J,J--Ks CMD of the central pointing of M15. The photometry was performed by 
using an asymmetric Moffat function. The photometry is very accurate well below the 
turn--off region ($J \sim$17.75 mag). Note that current CMD is based on instrumental 
magnitudes.
Right: Same as the left, but the photometry is based on a symmetric Moffat function. 
The quality of the photometry steadily decreases in the faint magnitude limit.}  
\end{figure*}

\section{Cluster Photometry}

We first performed the photometry on the stacked J- and Ks--band image. 
The master list includes stars with one measurement in the two bands. The 
same list was then used to perform the photometry on individual J, and 
Ks--band images. The final catalog includes stars with at least one 
measurement per band. 
%
Fig.~6 shows the NIR J, J-Ks CMD of the very crowded central pointing. 
The improvements in using an asymmetric analytical PSF fitting are quite 
clear. Data plotted in the left panel (asymmetric PSF) display well 
defined sequences along the Red Giant Branch (RGB, J--Ks=-0.7 mag), 
the Horizontal Branch (HB, Ks$\sim$14.5 mag, J--Ks=-1 mag), the sub giant branch 
(SGB, Ks$\sim$17 J--Ks=-0.75 mag) and in particular across the MSTO region 
(Ks$\sim$18, J--Ks=0.9 mag). On the other hand, the CMD based on 
a symmetric analytical PSF fitting shows broader evolved sequences 
(RGB, HB, SGB) and a steady decrease in photometric accuracy approaching 
the MSTO. 
This is a crucial difference since the precision in the evaluation of 
the absolute age of GCs is strongly affected by the photometric precision 
of the MSTO. Note that an uncertainty of the order of one tenth of a magnitude 
typically introduces in the comparison with cluster isochrones an uncertainty 
of the order of 1 Gyr. 

The above evidence suggests that the new reduction strategy devised to 
perform accurate analytical PSF photometry on NIR images collected with 
SCAO systems appears very promising. Current preliminary results appear  
even more interesting if we take into account the fact that the cluster 
regions covered by the central pointing are characterized by a very high 
central stellar density
Indeed, M15 belongs to the small sample of post core--collapsed GCs. 

\subsection{Absolute photometric calibration} 

The photometric catalog released by 2MASS has been a quantum jump in 
the improvement of the absolute calibration of NIR ground-based 
photometry. However, this catalog can be barely used to provide local 
standards to new AO systems available at the 8-m class telescopes. 
The reasons are manifold. The limiting magnitude of the 2MASS catalog 
is $J,H,Ks$=14--15 mag. This limit becomes even brighter in crowded 
stellar fields. These objects are typically saturated in NIR images 
collected with AO systems. The FOV of NIR detectors used with 
AO systems is at most of the order of 2$\times$2', this means 
that the number of local 2MASS standard stars is quite limited. 
It goes without saying that a precise calibration, 1--2 hundredths 
of magnitude, of NIR images collected with AO systems does require 
a sizable sample of local standards with apparent magnitudes of 
$J,H,Ks$=16--18 mag. To overcome this problem have been adopted 
both NIR images collected with both 4m and 8m telescopes (Bono et al. 
2010; Stetson et al. 2014).     

To calibrate NIR images collected with PISCES at LBT, we secured a 
set of NIR images collected with LUCI1 on June 2012. The key advantage 
of the LUCI1 images is that the LUCI1 FOV is 4$\times$4', thus it 
covers the two PISCES pointings. Data reduction for LUCI1 images was 
performed using the standard procedures DAOPHOT/ALLSTAR/ALLFRAME as described
in Monelli et al. 2010\cite{monelli10a}. The LUCI1 
NIR images were calibrated using 2MASS local standards. The NIR photometry 
based on LUCI1 images is very accurate down to $J,Ks \sim$19 mag and they 
were adopted to calibrate the photometric catalogs based on PISCES images
(Monelli et al. 2014, to be submitted). Finally, we mention that the use 
of a sizable sample of local standards overcomes the problem of the aperture 
correction.

  \begin{figure*}
   \begin{center}
   \begin{tabular}{c}
   \includegraphics[height=8cm]{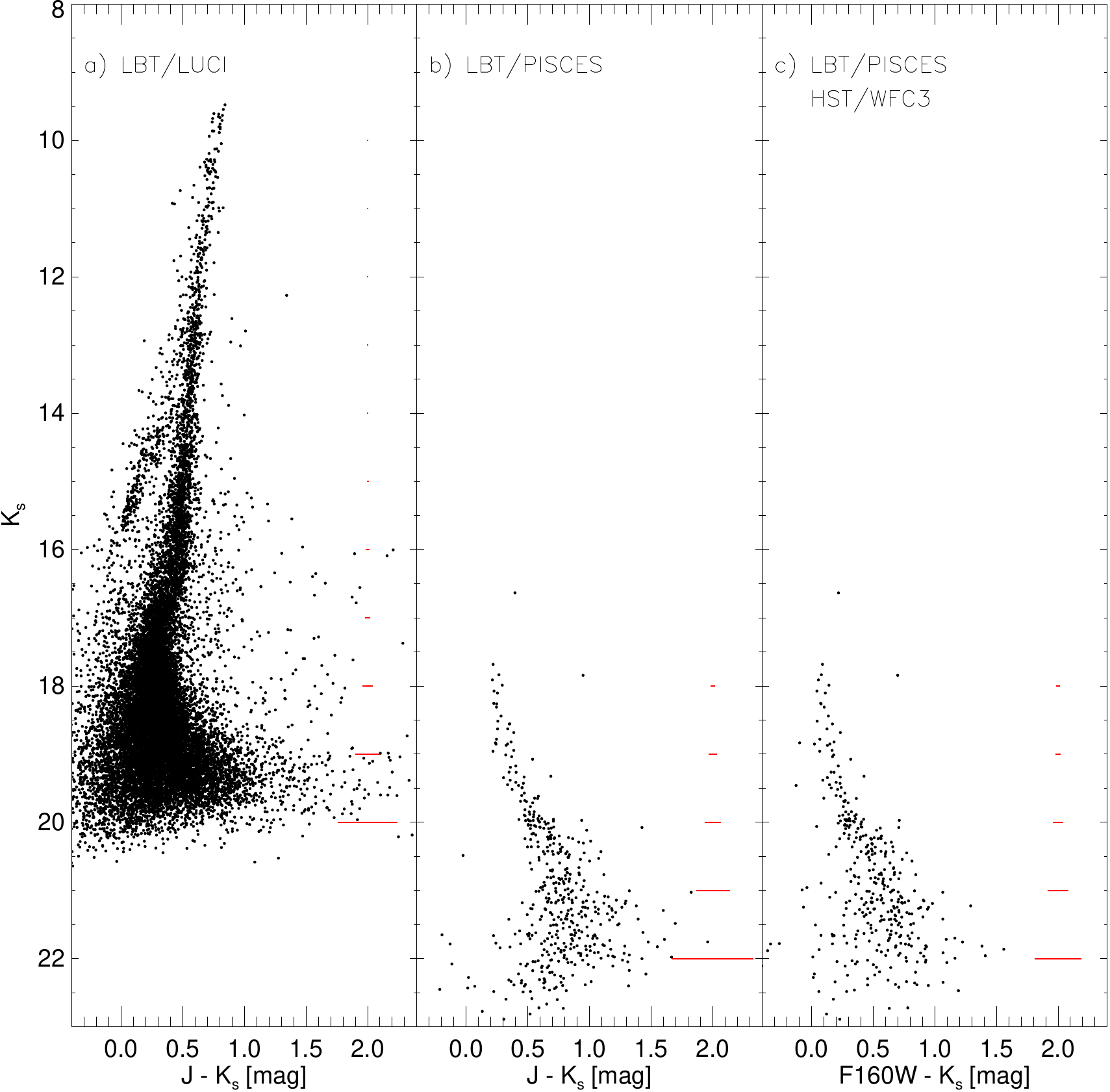}
   \end{tabular}
   \end{center}
   \caption[example] 
   { \label{fig7} 
Left: NIR Ks, J--Ks CMD based on seeing limited images collected with LUCI1 at 
LBT (left). The red error bars on the left display the error both in magnitude 
and in color.  Middle: Same as the left, but the photometry is based on images collected with PISCES at LBT with a SCAO (FLAO) system. Note that the limiting magnitude in the Ks--band is the deepest ever collected with a ground--based telescope. Right: Same as the left, but for the Ks, F160W-Ks CMD. The Fi60W photometry is based on space images collected with WFC3 at HST, while the Ks--band was collected with PISCES at LBT. Note that the F160W photometry is approximately one magnitude shallower.}
   \end{figure*}

\section{The Absolute Age of M15}
In this section we anticipate a new result detailed in Monelli et
al. (2014, to be submitted) concerning the absolute age of M15. This result
is based on data collected for the off--center pointing of M15.
To perform stellar photometry, we have used the same procedure
described above to analyse PISCES@LBT images: an updated
version of ROMAFOT with the new asymmetric PSF function (eq.~2). This resulted to be the best choice to obtain the
deepest CMD ever observed in near IR by reaching a Ks limiting
magnitude of $\sim$ 23 mag, see Fig.~\ref{fig7} (middle panel). 
In this figure we have also shown the Ks, J-Ks CMD obtained using the
camera LUCI1@LBT. As described above, these data were taken mainly for calibration
purpose and show the seeing--limited behavior of data taken
using the same telescope but a different camera with a pixel scale of
0.118''/px, about 10 times larger than PISCES'one. 

We have performed standard data reduction\cite{monelli10a} to analyse
space telescope data from WFC3@HST using F814W, F606W, and F160W bands. In this paper we show only the
results obtained with the near IR channel using F160W passband (see right panel in Fig.~\ref{fig7}) to make
a fair comparison between space and AO images taken from the ground with
LBT. The CMDs coming from the combination of the full set of optical and
near IR bands are presented in Monelli et al. 2014 (to be submitted).
The resulting near IR CMDs show clearly the jump in limiting magnitude
obtained when moving from seeing--limited IR collected with LUCI1 and
the AO assisted taken with PISCES@LBT. The reached limiting magnitude
is Ks$\sim$23 mag, that means about 3 mag fainter in Ks band when AO systems are used on the same
telescope. In spite of the small FOV corrected for atmosphere
turbulence, thanks the AO we are able for the first time to detect the
MSK in a very metal poor GGC. The quality obtained using J and Ks
bands with PISCES is very likely comparable to what is obtained using
a combination of space (F160W) and ground based (Ks) filters. However
the intrinsic errors given in J band are much larger than in F160W,
this is mainly due to the larger exposure time used for space
observation which is 3.5 times larger than what used to integrate
J band data.

In order to determine the absolute age of M15, we decided to use the
best photometry available, this means a combination of 
Ks--band from PISCES and optical (F606W) space observations. We have
used evolutionary tracks for an old age range (from 12 to 15
  Gyr, Vandenberg priv. comm.) with
[Fe/H]=-2.4 dex, an $\alpha-$enhancement $\alpha=$+0.4 and a
primordial helium abundance Y$=$0.25. These tracks were transformed into the
observational plane using colour--temperature relation provided by
Casagrande (priv. comm.). To match the data with isochrones we have adopted a distance
modulus of $\mu_0=$15.14 mag\cite{durrell93} and a reddening
E(B--V)$=$0.08\cite{sandage81c}.

At this point, we have all the ingredients to derive the absolute age
of M15. We approach at this estimation by using two independent
methods: the detection of the MSTO and the difference between the MSTO
and the MSK magnitudes, i.e. MSTO--MSK (fully described in 
Bono et al. 2010\cite{bono10b}). We notice that the first method does 
dependent on distance and reddening whereas the second one, being a 
differential method, is free from the uncertainties affecting these 
parameters. In fact, the distance modulus assumed for M15 is given 
with an error of $\pm$ 0.15 mag\cite{durrell93}. 

We have used CMD ridge lines to define both the MSTO and the MSK as
follows. The MSTO is identified as the bluer point in the TO
region. The MSK is defined as the maximum curvature point in the low
part of the MS. The same definition has been used to determine the
magnitude and colours of these two points for both data and
observations. In particular for near IR data, we have constrained the
MSTO and MSK locations using LUCI1 and PISCES data respectively. The evolutionary tracks allow us, at fixed chemical composition, to estimate the derivative of the MSTO magnitude as a function of age: $\Delta$F606W/$\Delta$t$=$ 0.083 mag/Gyr and $\Delta$Ks/$\Delta$t$=$ 0.051 mag/Gyr. The age corresponding to M15 is derived by interpolating the previous relation assuming the observed MSTO magnitude.
 The error budget has to take into account various sources. From the
 observational side, we included the error on the TO magnitude, the
 reddening and distance. In the case of the present data set, the
 photometric error varies depending on the filter used, from
 $\sim$0.015 in the case of the F606W filter to the $\sim$0.04 mag for
 the LUCI1 Ks. Using these photometric errors and the above
 derivatives we can estimate the absolute distance of M15 and its global
 error budget. The MSTO returns 13.97 $\pm$ 1.4
Gyr, which is in very good agreement with that obtained using the
MSTO--MSK method, i.e. 13.70$\pm$ 0.80
Gyr\footnote{The interested reader will find much more details in
  Monelli et al. (2014, to be submitted).}.

\section{Conclusions and final remarks}

We present new and accurate NIR (J,Ks) photometry for the Galactic globular cluster M15. 
The images were collected with PISCES at LBT with an innovative SCAO system equipped with 
high-order pyramid wave-front sensor (FLAO). The images were collected for two different 
pointings by using two natural guide stars with an apparent magnitude of R$\lsim$13 mag.     
The images were collected with medium seeing condition ($\approx$0.9 arcsec), but the 
images display a mean FWHM of 0.05 arcsec and of 0.06 arcsec in the J- and in the Ks--band 
images, respectively. The Strehl ratio on the quoted images ranges from 13--30\% (J) to 
50--65\% (Ks). Current results further support the evidence that new SCAO systems succeed 
in delivering high-quality NIR images in very crowded fields. However, the J-band 
images display a complex change in the shape of the PSF when moving at larger radial 
distances from the natural guide star. In particular, the stellar images become more 
elongated in approaching the corners of the J-band images. The Ks--band images are 
marginally affected by this limit.     

To overcome this thorny problem we devised a new reduction strategy based on asymmetric 
Moffat functions to perform accurate and deep photometry in crowded stellar fields. 
Current approach allowed us to reach in the off--center pointing a limiting magnitude of 
Ks$\sim$23 mag. This means, to our knowledge, the deepest K-band photometry ever collected 
with a ground--based telescope. The new NIR photometry and a well defined knee along the 
lower main sequence allowed us to estimate the absolute age of M15 with a precision 
that is a factor of two better when compared with similar estimates based on the MSTO 
available in the literature, i.e. using the MSTO--MSK we obtain 13.70$\pm$ 0.80
Gyr.    

Moreover, we used the photometry of the central pointing to constrain the precision 
of the new approach in performing accurate and photometry in crowded stellar fields. 
We found that asymmetric analytical PSF in the J-band images allows us to decrease 
by a factor of two the residuals in flux between the original and the reconstructed 
stellar profile. The difference in the Ks--band images is smaller and of the order of 
10\%. Current findings, once supported by independent investigations, appear very 
promising not only for constraining the absolute age of globular clusters, 
but also to improve current understanding of the stellar populations in the 
crowded regions of the Galactic center and of the Galactic Bulge. Even more importantly, the above results are opening new paths 
in the reduction and analysis of NIR images will be delivered by NIR detectors 
available at the next generation of Extremely Large Telescopes. These new 
observing facilities will be equipped with a challenging suite of AO systems 
(GLAO, SCAO, MCAO, MOAO). The development of new algorithms to perform 
accurate photometry seems a good viaticum to fully exploit the capabilities 
of these unprecedented observing facilities.

\acknowledgments{ 
GF has been supported by the Futuro in Ricerca di Base 2013 (RBFR13J716). This work
was partially supported by PRIN--INAF 2011 ``Tracing the formation
and evolution of the Galactic halo with VST'' (P.I.: M.~Marconi)
and by PRIN--MIUR (2010LY5N2T) ``Chemical and dynamical evolution
of the Milky Way and Local Group galaxies'' (P.I.: F.~Matteucci).} 




\bibliographystyle{spiebib}   

\begin{thebibliography}{35}

\bibitem{mcconnachie12}
{McConnachie}, A.~W., ``{The Observed Properties of Dwarf Galaxies in and
  around the Local Group},'' AJ~{\bf 144},  4 (July 2012).

\bibitem{marin09}
{Mar{\'{\i}}n-Franch}, A., {Aparicio}, A., {Piotto}, G., {Rosenberg}, A.,
  {Chaboyer}, B., {Sarajedini}, A., {Siegel}, M., {Anderson}, J., {Bedin},
  L.~R., {Dotter}, A., {Hempel}, M., {King}, I., {Majewski}, S., {Milone},
  A.~P., {Paust}, N., and {Reid}, I.~N., ``{The ACS Survey of Galactic Globular
  Clusters. VII. Relative Ages},'' ApJ~{\bf 694},  1498--1516 (Apr.
  2009).

\bibitem{leaman13}
{Leaman}, R., {VandenBerg}, D.~A., and {Mendel}, J.~T., ``{The bifurcated
  age-metallicity relation of Milky Way globular clusters and its implications
  for the accretion history of the galaxy},'' MNRAS~{\bf 436},  122--135
  (Nov. 2013).

\bibitem{kasper10}
{Kasper}, M., {Beuzit}, J.-L., {Verinaud}, C., {Gratton}, R.~G., {Kerber}, F.,
  {Yaitskova}, N., {Boccaletti}, A., {Thatte}, N., {Schmid}, H.~M., {Keller},
  C., {Baudoz}, P., {Abe}, L., {Aller-Carpentier}, E., {Antichi}, J.,
  {Bonavita}, M., {Dohlen}, K., {Fedrigo}, E., {Hanenburg}, H., {Hubin}, N.,
  {Jager}, R., {Korkiakoski}, V., {Martinez}, P., {Mesa}, D., {Preis}, O.,
  {Rabou}, P., {Roelfsema}, R., {Salter}, G., {Tecza}, M., and {Venema}, L.,
  ``{EPICS: direct imaging of exoplanets with the E-ELT},'' in [{\em Society of
  Photo-Optical Instrumentation Engineers (SPIE) Conference
  Series}{\nolinebreak\hspace{0.1em}]},  {\em Society of Photo-Optical
  Instrumentation Engineers (SPIE) Conference Series} {\bf 7735} (July 2010).

\bibitem{deep11}
{Deep}, A., {Fiorentino}, G., {Tolstoy}, E., {Diolaiti}, E., {Bellazzini}, M.,
  {Ciliegi}, P., {Davies}, R.~I., and {Conan}, J.-M., ``{An E-ELT case study:
  colour-magnitude diagrams of an old galaxy in the Virgo cluster},'' A\&A~{\bf 531},  A151 (July 2011).

\bibitem{greggio12}
{Greggio}, L., {Falomo}, R., {Zaggia}, S., {Fantinel}, D., and {Uslenghi}, M.,
  ``{Resolved Stellar Population of Distant Galaxies in the ELT Era},'' PASP~{\bf 124},  653--667 (July 2012).

\bibitem{schreiber13}
{Schreiber}, L., {La Camera}, A., {Prato}, M., {Diolaiti}, E., {constraint on
  the PSFS which is an upper bound derived from the Strehl ratio (SR)}, h. s.
  i. g. w. s. P. a. f. t. P.-A. s. o. t. E.-E. M. s.~M., and different~crowding
  conditions., ``{Point Spread Function extraction in crowded fields using
  blind deconvolution},'' in [{\em Proceedings of the Third AO4ELT
  Conference}{\nolinebreak\hspace{0.1em}]},  {Esposito}, S. and {Fini}, L.,
  eds. (Dec. 2013).

\bibitem{marchetti08}
{Marchetti}, E., {Brast}, R., {Delabre}, B., {Donaldson}, R., {Fedrigo}, E.,
  {Frank}, C., {Hubin}, N., {Kolb}, J., {Lizon}, J.-L., {Marchesi}, M.,
  {Oberti}, S., {Reiss}, R., {Soenke}, C., {Tordo}, S., {Baruffolo}, A.,
  {Bagnara}, P., {Amorim}, A., and {Lima}, J., ``{MAD on sky results in star
  oriented mode},'' in [{\em Society of Photo-Optical Instrumentation Engineers
  (SPIE) Conference Series}{\nolinebreak\hspace{0.1em}]},  {\em Society of
  Photo-Optical Instrumentation Engineers (SPIE) Conference Series} {\bf 7015}
  (July 2008).

\bibitem{momany08}
{Momany}, Y., {Ortolani}, S., {Bonatto}, C., {Bica}, E., and {Barbuy}, B.,
  ``{Multi-Conjugate Adaptive Optics VLT imaging of the distant old open
  cluster FSR1415},'' MNRAS~{\bf 391},  1650--1658 (Dec. 2008).

\bibitem{gullieuszik08}
{Gullieuszik}, M., {Greggio}, L., {Held}, E.~V., {Moretti}, A., {Arcidiacono},
  C., {Bagnara}, P., {Baruffolo}, A., {Diolaiti}, E., {Falomo}, R., {Farinato},
  J., {Lombini}, M., {Ragazzoni}, R., {Brast}, R., {Donaldson}, R., {Kolb}, J.,
  {Marchetti}, E., and {Tordo}, S., ``{Resolving stellar populations outside
  the Local Group: MAD observations of UKS 2323-326},'' A\&A~{\bf 483},
  L5--L8 (May 2008).

\bibitem{moretti09}
{Moretti}, M.~I., {Dall'Ora}, M., {Ripepi}, V., {Clementini}, G., {Di
  Fabrizio}, L., {Smith}, H.~A., {DeLee}, N., {Kuehn}, C., {Catelan}, M.,
  {Marconi}, M., {Musella}, I., {Beers}, T.~C., and {Kinemuchi}, K., ``{The Leo
  IV Dwarf Spheroidal Galaxy: Color-Magnitude Diagram and Pulsating Stars},''
  ApJL~{\bf 699},  L125--L129 (July 2009).

\bibitem{bono10b}
{Bono}, G., {Caputo}, F., {Marconi}, M., and {Musella}, I., ``{Insights into
  the Cepheid Distance Scale},'' ApJ~{\bf 715},  277--291 (May 2010).

\bibitem{fiorentino11}
{Fiorentino}, G., {Tolstoy}, E., {Diolaiti}, E., {Valenti}, E., {Cignoni}, M.,
  and {Mackey}, A.~D., ``{MAD about the Large Magellanic Cloud. Preparing for
  the era of Extremely Large Telescopes},'' A\&A~{\bf 535},  A63 (Nov.
  2011).

\bibitem{ferraro09a}
{Ferraro}, F.~R., {Dalessandro}, E., {Mucciarelli}, A., {Beccari}, G., {Rich},
  R.~M., {Origlia}, L., {Lanzoni}, B., {Rood}, R.~T., {Valenti}, E.,
  {Bellazzini}, M., {Ransom}, S.~M., and {Cocozza}, G., ``{The cluster Terzan 5
  as a remnant of a primordial building block of the Galactic bulge},'' Nature~{\bf 462},  483--486 (Nov. 2009).

\bibitem{esposito11}
{Esposito}, S., {Riccardi}, A., {Pinna}, E., {Puglisi}, A.,
  {Quir{\'o}s-Pacheco}, F., {Arcidiacono}, C., {Xompero}, M., {Briguglio}, R.,
  {Agapito}, G., {Busoni}, L., {Fini}, L., {Argomedo}, J., {Gherardi}, A.,
  {Brusa}, G., {Miller}, D., {Guerra}, J.~C., {Stefanini}, P., and {Salinari},
  P., ``{Large Binocular Telescope Adaptive Optics System: new achievements and
  perspectives in adaptive optics},'' in [{\em Society of Photo-Optical
  Instrumentation Engineers (SPIE) Conference
  Series}{\nolinebreak\hspace{0.1em}]},  {\em Society of Photo-Optical
  Instrumentation Engineers (SPIE) Conference Series} {\bf 8149} (Sept. 2011).

\bibitem{allard12}
{Allard}, F., {Homeier}, D., and {Freytag}, B., ``{Stellar to Substellar Model
  Atmospheres},'' in [{\em IAU Symposium}{\nolinebreak\hspace{0.1em}]},
  {Richards}, M.~T. and {Hubeny}, I., eds., {\em IAU Symposium} {\bf 282},
  235--242 (Apr. 2012).

\bibitem{allard13}
{Allard}, F., {Homeier}, D., {Freytag}, B., {Schaffenberger}, {}, W., and
  {Rajpurohit}, A.~S., ``{Progress in modeling very low mass stars, brown
  dwarfs, and planetary mass objects.},'' {\em Memorie della Societa
  Astronomica Italiana Supplementi}~{\bf 24},  128 (2013).

\bibitem{king71}
{King}, I.~R., ``{The Profile of a Star Image},'' PASP~{\bf 83},  199
  (Apr. 1971).

\bibitem{vandermarel02}
{van der Marel}, R.~P., {Gerssen}, J., {Guhathakurta}, P., {Peterson}, R.~C.,
  and {Gebhardt}, K., ``{Hubble Space Telescope Evidence for an
  Intermediate-Mass Black Hole in the Globular Cluster M15. I. STIS
  Spectroscopy and WFPC2 Photometry},'' AJ~{\bf 124},  3255--3269 (Dec.
  2002).

\bibitem{stetson94}
{Stetson}, P.~B., ``{The center of the core-cusp globular cluster M15: CFHT and
  HST Observations, ALLFRAME reductions},'' PASP~{\bf 106},  250--280
  (Mar. 1994).

\bibitem{stetson87}
{Stetson}, P.~B., ``{DAOPHOT - A computer program for crowded-field stellar
  photometry},'' PASP~{\bf 99},  191--222 (Mar. 1987).

\bibitem{stetson90}
{Stetson}, P.~B., ``{On the growth-curve method for calibrating stellar
  photometry with CCDs},'' PASP~{\bf 102},  932--948 (Aug. 1990).

\bibitem{buonanno83}
{Buonanno}, R., {Buscema}, G., {Corsi}, C.~E., {Ferraro}, I., and {Iannicola},
  G., ``{Automated photographic photometry of stars in globular clusters},''
  A\&A~{\bf 126},  278--282 (Oct. 1983).

\bibitem{buonanno89}
{Buonanno}, R. and {Iannicola}, G., ``{Stellar photometry with big pixels},''
  PASP~{\bf 101},  294--301 (Mar. 1989).

\bibitem{schechter93}
{Schechter}, P.~L., {Mateo}, M., and {Saha}, A., ``{DOPHOT, a CCD photometry
  program: Description and tests},'' PASP~{\bf 105},  1342--1353 (Nov.
  1993).

\bibitem{bertin96}
{Bertin}, E. and {Arnouts}, S., ``{SExtractor: Software for source
  extraction.},'' A\&AS~{\bf 117},  393--404 (June 1996).

\bibitem{bertin11}
{Bertin}, E., ``{Automated Morphometry with SExtractor and PSFEx},'' in [{\em
  Astronomical Data Analysis Software and Systems
  XX}{\nolinebreak\hspace{0.1em}]},  {Evans}, I.~N., {Accomazzi}, A., {Mink},
  D.~J., and {Rots}, A.~H., eds., {\em Astronomical Society of the Pacific
  Conference Series} {\bf 442},  435 (July 2011).

\bibitem{diolaiti00}
{Diolaiti}, E., {Bendinelli}, O., {Bonaccini}, D., {Close}, L., {Currie}, D.,
  and {Parmeggiani}, G., ``{Analysis of isoplanatic high resolution stellar
  fields by the StarFinder code},'' A\&AS~{\bf 147},  335--346 (Dec.
  2000).

\bibitem{anderson06}
{Anderson}, J. and {King}, I.~R., ``{PSFs, Photometry, and Astronomy for the
  ACS/WFC},'' tech. rep. (Feb. 2006).

\bibitem{anderson10}
{Anderson}, J. and {van der Marel}, R.~P., ``{New Limits on an
  Intermediate-Mass Black Hole in Omega Centauri. I. Hubble Space Telescope
  Photometry and Proper Motions},'' ApJ~{\bf 710},  1032--1062 (Feb.
  2010).

\bibitem{bedin09}
{Bedin}, L.~R., {Salaris}, M., {Piotto}, G., {Anderson}, J., {King}, I.~R., and
  {Cassisi}, S., ``{The End of the White Dwarf Cooling Sequence in M4: An
  Efficient Approach},'' ApJ~{\bf 697},  965--979 (June 2009).

\bibitem{lagioia14}
{Lagioia}, E.~P., {Milone}, A.~P., {Stetson}, P.~B., {Bono}, G., {Prada
  Moroni}, P.~G., {Dall'Ora}, M., {Aparicio}, A., {Buonanno}, R., {Calamida},
  A., {Ferraro}, I., {Gilmozzi}, R., {Iannicola}, G., {Matsunaga}, N.,
  {Monelli}, M., and {Walker}, A., ``{On the Kinematic Separation of Field and
  Cluster Stars across the Bulge Globular NGC 6528},'' ApJ~{\bf 782},
  50 (Feb. 2014).

\bibitem{monelli10a}
{Monelli}, M., {Hidalgo}, S.~L., {Stetson}, P.~B., {Aparicio}, A., {Gallart},
  C., {Dolphin}, A.~E., {Cole}, A.~A., {Weisz}, D.~R., {Skillman}, E.~D.,
  {Bernard}, E.~J., {Mayer}, L., {Navarro}, J.~F., {Cassisi}, S., {Drozdovsky},
  I., and {Tolstoy}, E., ``{The ACS LCID Project. III. The Star Formation
  History of the Cetus dSph Galaxy: A Post-reionization Fossil},'' ApJ~{\bf 720},  1225--1245 (Sept. 2010).

\bibitem{durrell93}
{Durrell}, P.~R. and {Harris}, W.~E., ``{A color-magnitude study of the
  globular cluster M15},'' AJ~{\bf 105},  1420--1440 (Apr. 1993).

\bibitem{sandage81c}
{Sandage}, A., {Katem}, B., and {Sandage}, M., ``{The Oosterhoff period groups
  and the age of globular clusters. I Photometry of cluster variables in
  M15},'' ApJS~{\bf 46},  41--74 (May 1981).

\end{thebibliography}

\end{document}